\documentclass[]{raa}
\usepackage{graphicx,times}
\input{psfig.sty}
\input{epsf.sty}
\begin{document}
\title{The Deep Optical Imaging of the Extended Groth Strip}
\volnopage{Vol.0 (200x) No.0, 000--000}
\setcounter{page}{1} 
\author{
Ying-He Zhao\inst{1,2,3}
\and Jia-Sheng Huang\inst{2}
\and M. L. N.  Ashby\inst{2}
\and G. G.  Fazio\inst{2}
\and S. Miyazaki\inst{4}}
\institute{Purple Mountain Observatory, Chinese Academy of Sciences (CAS), Nanjing 210008, China;
{\it yhzhao@pmo.ac.cn}\\
\and Harvard-Smithsonian Center for Astrophysics, 60 Garden Street, Cambridge, MA 02138, USA;
\and Department of Astronomy, Nanjing University, Nanjing 210093, China
\and National Astronomical Observatory of Japan, Mitaka, Tokyo 181-8588, Japan}

\date{Received~~2009 month day; accepted~~2009~~month day}

\abstract{
We present $u'g'R$ optical images taken with the MMT/Megacam and the Subaru/Suprime of the Extended Groth
Strip survey. The total survey covers an area of about $\sim 1$ degree$^2$, including four sub-fields and is
optimized for the study of galaxies at $z\sim3$. Our methods for photometric calibration in AB magnitudes, the
limiting magnitude and the galaxy number count are described. A sample of 1642 photometrically selected
candidate LBGs to an apparent $R_{AB}$ magnitude limit of 25.0 is present. The average sky surface density of 
our LBGs sample is $\sim$ 1.0 arcmin$^{-2}$, slightly higher than the previous finding.
}

\authorrunning{Y.-H. Zhao et al.}
\titlerunning{The Deep Optical Imaging of the EGS}
\maketitle
\section{Introduction}
\label{sect:intro}

When and how galaxies formed is one of the primary questions in astronomy today. Observations of young
galaxies at high redshift are straightforward approach to this problem. Taking advantage of the progress in
observing technology and also sophistication of selection methods, (e.g., color selection, Steidel \&
Hamilton 1993; Steidel et al. 2003; Franx et al. 2004; Daddi et al. 2004), multicolor surveys allow for the
location of high redshift galaxies. Among the various methods the Lyman break dropout technique (Steidel \&
Hamilton 1993), sensitive to the presence of the 912 \AA\  break, is designed to select z $\sim $ 3 galaxies;
and it has been successfully used to find many young star-forming galaxies beyond z $\sim $ 2 (e.g., Steidel
et al. 2003; Lehnert \& Bremer 2003; Ouchi et al. 2004; Dickinson et al. 2004; Sawicki \& Tompson 2006).

Since their first discovery in the 1990s, various properties of Lyman Break Galaxies (LBGs) have been
extensively studied (e.g. the luminosity function and the clustering property). However, the stellar mass,
which is a robust tool to probe galaxy evolution, has been estimated only based on ground-based photometry.
For high redshift galaxies, this only samples out to rest-frame optical band. The observed luminosity at
these wavelengths is dominated by recent star-formation activity rather than the stellar population that has
accumulated over the galaxy's lifetime. With the advent of the Spitzer Space Telescope (hereafter Spitzer,
Werner et al. 2004), we can probe rest-frame near-infrared (NIR; $z$, J, H, K) band (for z$\sim 3$ galaxies) where the light is more sensitive to the total stellar mass, which is dominated by the lower mass stars. In addition, the longer wavelengths are less affected by dust extinction that may strongly attenuate ultraviolet light.  Shapley et al. (2005) showed that the addition of Spitzer IRAC data reduces stellar mass uncertainties by a factor of $1.5-2$ relative to estimates based on optical-$K_s$ photometry alone, for their LBGs sample of $z\sim2$. Elsner et al. (2008) found that the stellar mass is to be overestimated by more than a factor of three on average for $z\sim3$ galaxies without Spitzer data. 

Huang et al. (2005) has shown that, at a limiting flux densities (5$\sigma$) of 0.5 $\mu$Jy at 3.6 and 4.5 $\mu$m, 2.7 $\mu$Jy at 5.8 and 8.0 $\mu$m, the detection rates for z$\sim$3 LBGs are about 83\%, 78\%, 21\% and 17\%, at 3.6, 4.5, 5.8 and 8.0 $\mu$m, respectively. Preliminary results of adding IRAC photometry to stellar mass estimates have been presented in e.g. Barmby et al. (2004), Rigopoulou et al. (2006) and Huang et al. (2007) for z$\sim 3$ LBGs.

Despite of the stellar mass of LBGs, we can also use the Spitzer deep mid- and far-infrared (MIPS; Rieke et al. 2004) data study the link between UV-selected LBGs ($\sim$5\% detections in MIPS imaging data, Huang et al. 2005) and infrared-selected Ultra Luminous Infrared Galaxies (ULIRGs) at high redshift. To take advantage of the deep Spitzer IRAC+MIPS imaging data to study the physical properties and evolutionary processes of galaxies at redshift $\geq 3$, we began our optical imaging of the Extended Groth Strip (EGS: $\alpha$=14$^h$17$^m$, $\delta$=+52$^\circ$30$'$) sky field using the MEGACAM on the MMT (\emph{u'-} and \emph{g'-}band) and the Subaru telescope (\emph{R}-band). The EGS region is an extension of a \emph{Hubble Space Telescope (HST)} survey, which is consisting of 28 Wide-Field Planetary Camera 2 (WFPC2) pointings carried out in 1994 by the WFPC team (Rhodes et al. 2000). This field benefits from low extinction, low Galactic and zodiacal infrared emission and has attracted a wide range of deep observations at essentially every accessible wavelength (e.g. the All-Wavelength Extended Groth Strip 
International Survey, AEGIS; Davis et al. 2007). Although the ongoing Canada-France-Hawaii Telescope Legacy Survey (CFHTLS)\footnotemark\footnotetext{See http://www1.cadc-ccda.hia-iha.nrc-cnrc.gc.ca/community/CFHTLS-SG/docs/cfhtls.html} has carried out optical imaging observations, there are three things that we need improve  (1) our $u'$ filter cuts at 4000 \AA\ on the red side , which is about 200 \AA\ blueward of their $u'$ filter. The $u'/g'$ filter combination is optimally placed for measuring the Lyman continuum break in galaxies at $z$$\sim$3;  (2) their survey only covers about half area of the Spitzer data set; and (3) their observations are not very deep (as in Section 3.1). 

This paper is organized as following, in section 2, we present a brief description for the observations and 
data pre-reductions. We also describe the method of photometric calibration (eg magnitude zeropoints, the 
estimate of the sensitivity of the mosaics) in this section. We present the results of galaxy number counts 
at $u'g'R$ bands, and present a photometric selected LBG sample at $z$$\sim$3 in section 3. In the last 
section we present a summary for this paper.

\section{Observation and Data Reduction}
\label{sect:Obs}
\begin{table}[pbht]
\begin{center}
\caption[]{Journal of optical observations of EGS}
\label{table1}
\begin{tabular}{ccccccc}
\hline\noalign{\smallskip}
&$\alpha$ &$\delta$&& &Seeing&Zero point\\
Field &(J2000.0)$^a$&(J2000.0)$^a$&Filter&Telescope&(arcsec)&(mag)\\
\hline\noalign{\smallskip}
0 &14 16 00.84&+52 11 59.1&$u'$&MMT&0.84&31.01\\
& & &$g'$&MMT&1.0&33.53\\
& & &$R$&Subaru&0.70&31.923\\
1 &14 18 07.85&+52:35:34.8&$u'$&MMT&1.08&30.95\\
& & &$g'$&MMT&1.33&33.50\\
& & &$R$&Subaru&0.60&31.881 \\
2 &14 20 18.44&+52 59 11.0&$u'$&MMT&0.95&31.44\\
& & &$g'$&MMT&1.41&33.83\\
& & &$R$&Subaru&0.70&31.860 \\
3 &14 22 30.83&+53 22 51.0&$u'$&MMT&0.88&31.13\\
& & &$g'$&MMT&1.60&34.57\\
& & &$R$&Subaru&0.72&31.911\\
\hline\noalign{\smallskip}
\end{tabular}
\end{center}
\tablenotes{a}{0.72\textwidth}{Postions of the field centers. Units of right ascension are hours, minutes and seconds, and units of declination are degrees, arcminutes and arcseconds.}
\end{table}

\begin{figure}[pbht]
\begin{center}
\includegraphics[width=0.8\textwidth]{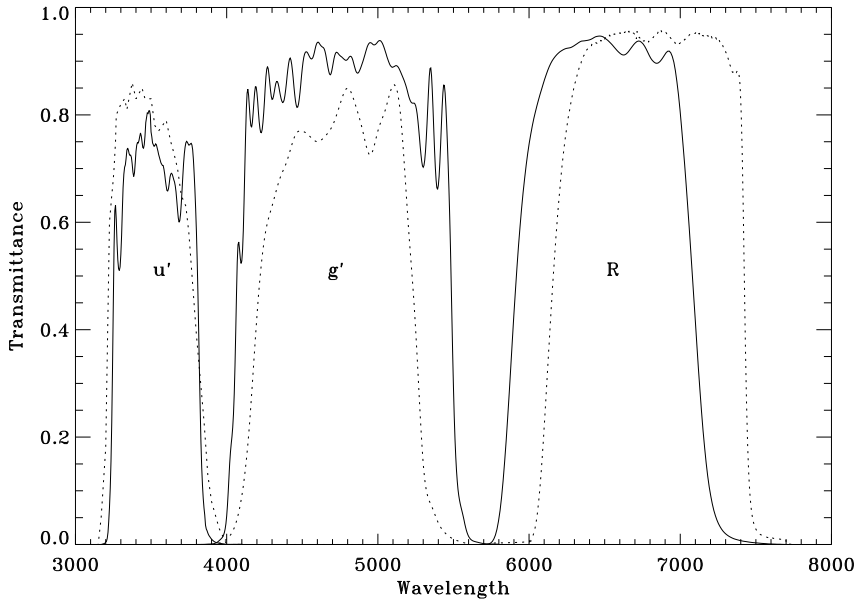} 
\caption{Filter system used for the deep imaging, $u'$ (3500/600), $g'$ (4750/1500) and $R$ (6500/1200). The
dashed line shows the filter system which is used by Steidel et al. (2003).}
\end{center}
\end{figure}

The imaging data for these four sub-fields presented here were obtained on the MMT and Subaru telescopes
during 2005, detailed in Table 1. The imaging filters used were the SDSS $u'g'$ and the Johnson-Cousins $R$
system. The filter system passbands are shown in figure 1, with the $U_nG\mathcal{R}$ system (Steidel et al.
2003) superposed. Our $u'$ filter is identical to the $U_n$ filter. The $g'$ filter has similar effective
wavelength as $G$ but is wider. The $R$ filter is shorter and slightly narrower than the $\mathcal{R}$
filter. This is the main cause of the difference of the source colors between these two filter systems.

In general, individual integrations of 340-500 ($u'$) and 600 or 1200 ($g'$) seconds were
obtained by using the MMT/Megacam with a pixel scale of 0.08$''$. The $R$-band images were conducted by the
Subaru/Suprime-Cam (Miyazaki et al. 2002) with a pixel scale of 0.202$''$. Total integration times varied depending on observing conditions, filters and
telescope/camera combinations.

The data reduction for the $u'$- and $g'$-band images, such as doing bias and flat corrections, mosaics,
flux and astrometrical calibrations, were carried out by Ashby, and we will only give a brief description
here. The details of these procedures can be found at Ashby's homepage\footnotemark
\footnotetext{http://www.cfa.harvard.edu/$\sim$mashby/megacam/megacam\_frames.html}. 
While for the $R$-band images, please refer to Miyazaki et al. (2002) and Miyazaki et al. (2007) for the detailed
data reduction procedures.

In brief, after CCD processing, the images were astrometrically calibrated with reference to the USNO-B1.0
astrometric catalog (Monet et al. 2003), using polynomial solution to map the focal plane to the astrometric
reference. The resulting accuracy is better than 0.2$''$.

The twilight flats at $u'$, $g'$ are somewhat contaminated by OH skyglow emission that can sneak in to
varying degrees depending on distance from the center of the megacam FOV. Therefore, we need correct for the
OH contribution to the twilight flat before flux calibrations. The OH contamination was
compensated for by observing a part of our field (F1) during photometric
conditions at a number of positions with short exposures (60s) intended to
match the depth of the SDSS survey.  We then used the multiple
observations of stars in those short integrations to correct the
contaminated flat by measuring the variation in apparent brightness of
those stars with respect to their distance from the MMT boresight.  The
corrections thereby derived were applied to all individual Megacam frames
before we constructed our coadds. Then we generated the
catalogs by running {\scriptsize SExtractor} (Bertin \& Arnouts 1996) on the images, and then computed the
zeropoint for each image by position-matching these catalogs to the SDSS (Gunn et al. 1998) catalogs for the
same field. In Table 1 we also list the resulting magnitude zero points for these images. The magnitudes in this
paper are all given in AB magnitudes (Oke \& Gunn 1983; Fukugita et al. 1996).

The calibrated images were coadded using the {\scriptsize SWarp} written by E.~ Bertin of Terapix, and then
were combined as full mosaics using the IRAF\footnotemark \footnotetext{IRAF is distributed by the National
Optical Astronomy Observatory, which is operated by the Association of Universities for Research in
Astronomy, Inc., under cooperative agreement with the National Science Foundation.} task $imcombine$. The
mosaics of $u'$- and $g'$-band images have a pixel scale of 0.318$''$, larger than that of the $R$-band
images. We rebinned and rotated the $R$-band image to match the $u'$- and $g'$-band images, in order to
obtain the color information of each source in the image by using {\scriptsize SExtractor}'s double-image
mode. The process of rebinning will cause a change of -0.98 [$\equiv -$2.5log$_{10}(0.318/0.202)^2$] mag in the $R$
magnitude zero point.

In order to have a good estimate of the observational qualities for each field and filter, we run
{\scriptsize SExtractor version 2.5} on each image individually, not the rebinned image, to generate the
photometric catalogue. The magnitudes used for the further analyses are all measured in Kron-style apertures
({\scriptsize SExtractor}'s MAG\_AUTO), except for the section of LBG selections. These should give good
estimates of the total magnitudes for each sources, especially for the stars.

Stellar color is a useful diagnostic to examine the photometric magnitude zeropoint. Stars occupy a
relatively constrained locus in the color-color space. Any offsets between the observed and synthetic colors
indicates a zeropoint error. This test can be applied to our four fields, which lie in the SDSS (Gunn et al.
1998) field.

\subsection{Stellar Color-Color Diagram}
\begin{figure}[pbht]
\begin{center} 
\includegraphics[width=0.8\textwidth,bb=0 18 483 350]{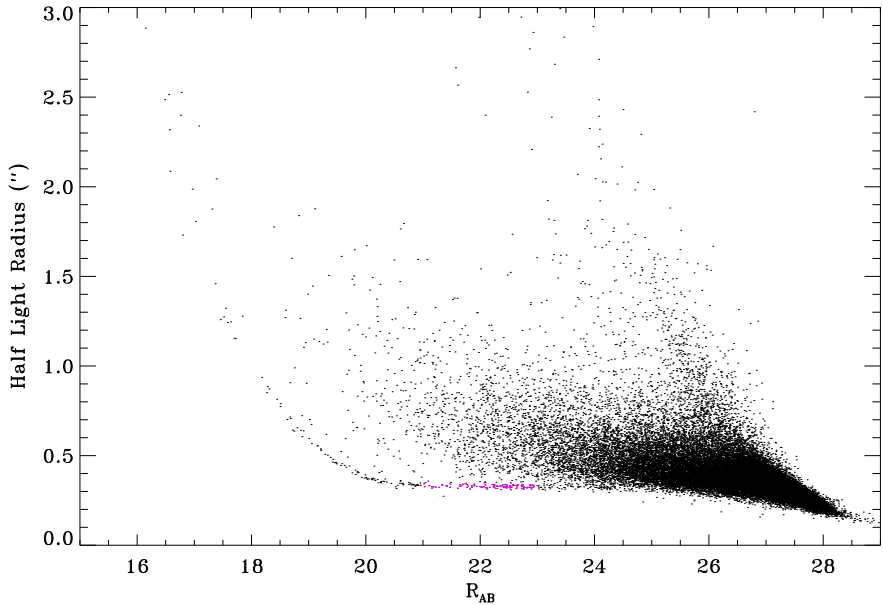}
\caption{Half-light radius plotted versus total magnitude. The horizontal branch shows locus of the stars.
The upturn at the bright end indicates the sources are saturated. The pink points are the conservative cuts
in magnitude and radius to select stars for further analysis. }
\end{center}
\end{figure}

\begin{figure}[pbht]
\begin{center} 
\includegraphics[width=0.77\textwidth,bb=22 8 446 321]{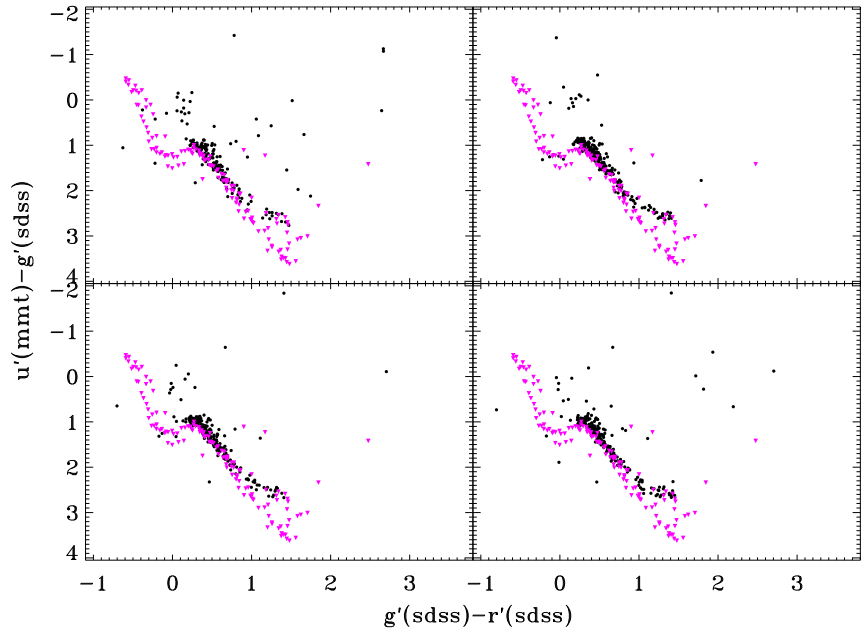}
\caption{The color-color diagram for $u'$-band images. The four panels are corresponding to the four
different fields. \emph{Upper left}: Field 0; \emph{Upper right}: Field 1; \emph{Lower left}: Field 2;
\emph{Lower right}: Field 3. We can see that it is very consistent between the observed (black circles) and
synthetic (magenta triangles) colors, which indicates that the magnitude zero points are correct.}
\end{center}
\end{figure}
\begin{figure}[pbht]
\begin{center} 
\includegraphics[width=0.77\textwidth,bb=22 8 446 321]{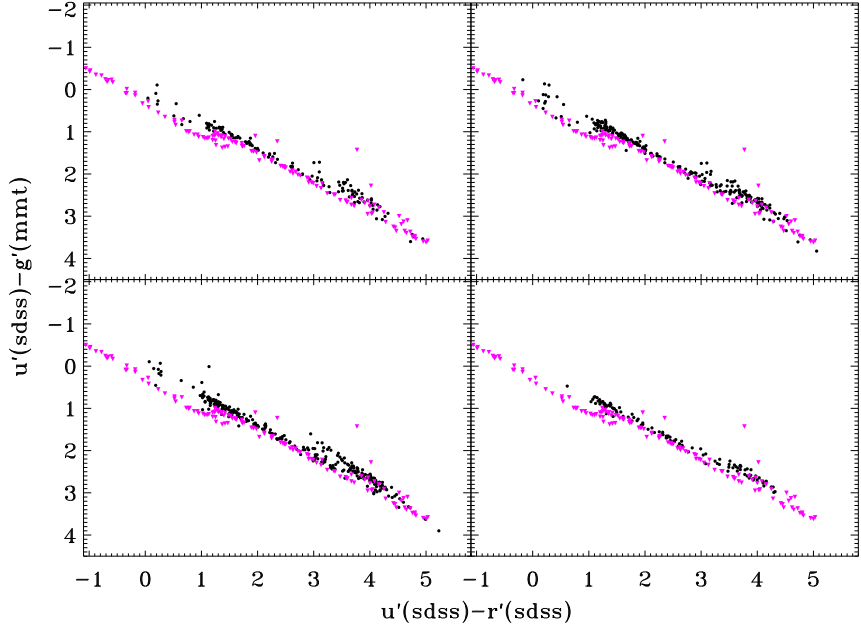}
\caption{Same as Figure 2, but for $g'$-band images. The small systemic offsets of $\sim 0.05$ mag on
$(u'-g')$ are revealed. }
\end{center}
\end{figure}
\begin{figure}[pbht]
\begin{center} 
\includegraphics[width=0.8\textwidth,bb=22 8 446 321]{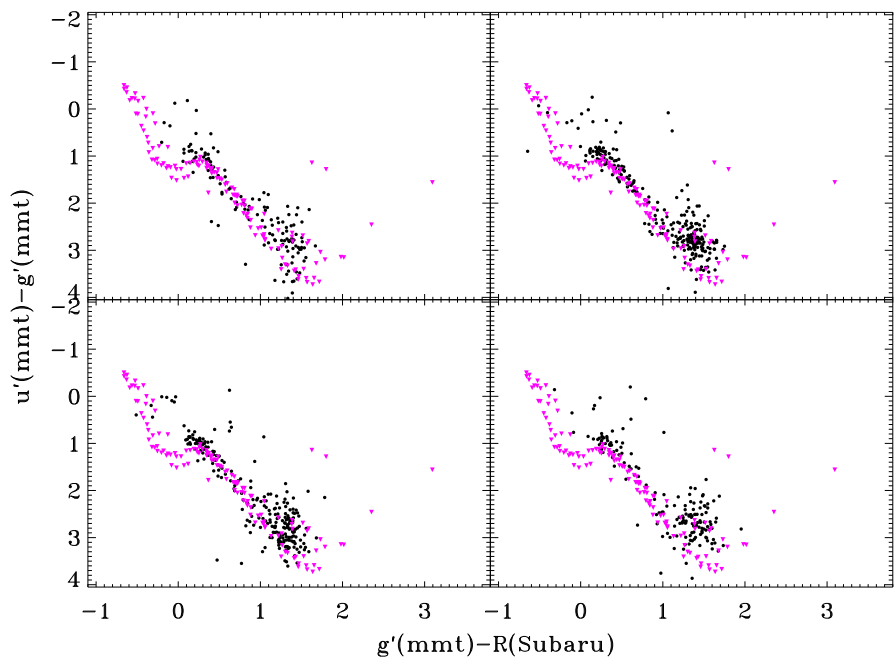}
\caption{Same as figure 2, for $R$-band images. Although there exists larger dispersion than the other two
band figures, there are no obvious offsets between the observed and synthetic colors. We have corrected the
zero point errors for the $g'$-band magnitudes when plotting this figure.}
\end{center}
\end{figure}

\begin{figure}[pbht]
\begin{center} 
\includegraphics[width=0.8\textwidth,bb=22 8 446 321]{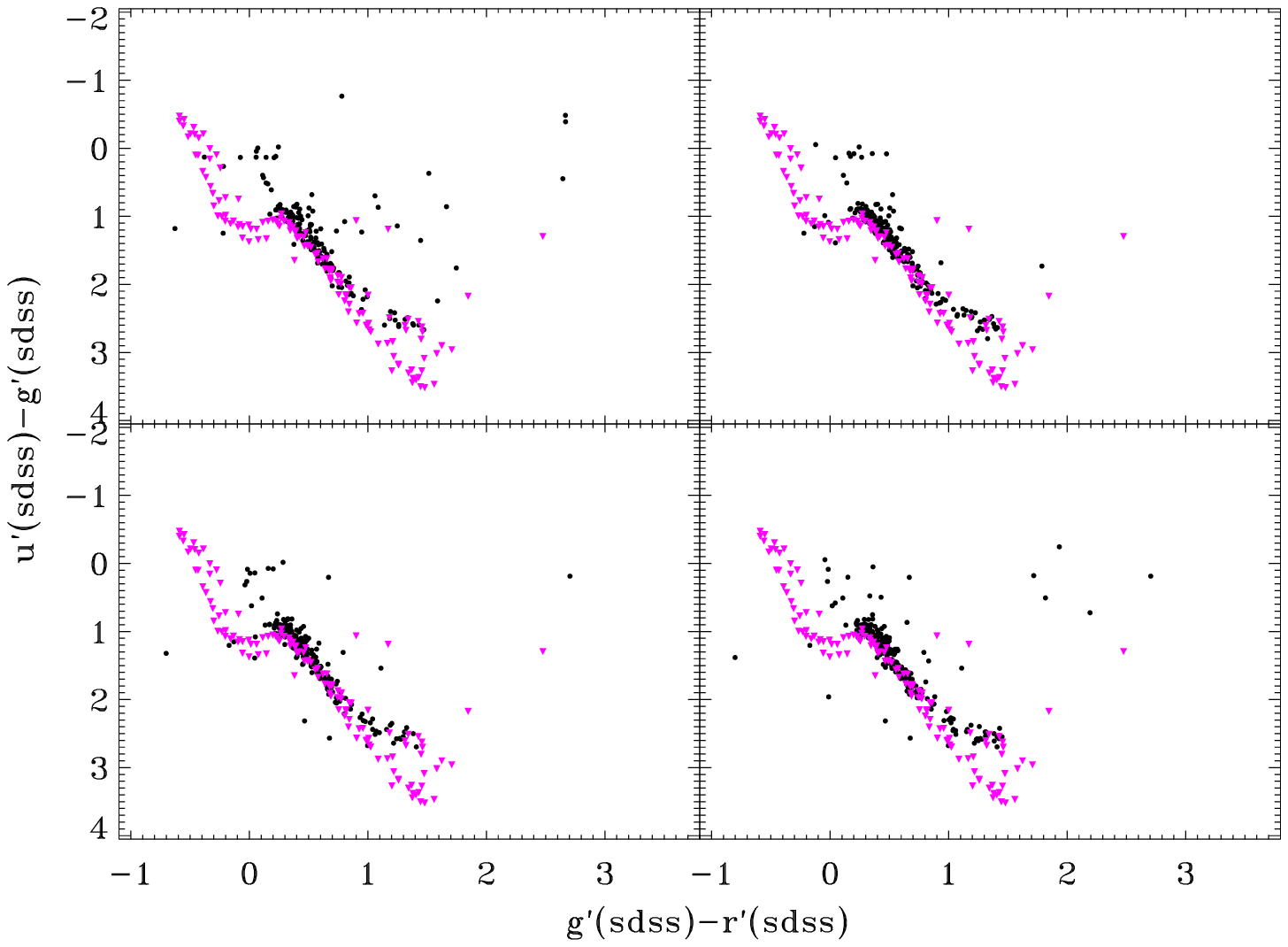} 
\caption{Same as figure 2, for SDSS $u'g'r'$ filters.}
\end{center}
\end{figure}

In Figure 2 we illustrates the selection of stars. The plot shows half-light radius plotted versus
magnitude. In this figure, the stars show up a well defined horizontal locus while the galaxies occupy a
range of magnitudes and radii. The upturn at the bright end indicates the sources are saturated. The pink
points indicate the cuts in magnitude and radius to select stars for the further analysis.

Figure 3, 4 and 5 show color-color diagrams for stars both in the EGS and SDSS fields, for $u'$-, $g'$- and
$R$-bands, respectively. The magenta triangles and solid circles in each figure represent synthetic and
observed colors, respectively. The different panels in each figure are different fields. The synthetic
colors are obtained using the observed stellar spectra library HILIB (Pickles 1998) to be convolved with
the transmission curve of the filters. Since we do not get the response curve, we do not apply the CCD
quantum efficiency (QE) for MMT/Subaru filters during this calculation. However, the results of synthetic
colors show good agreement with the observed colors, except for a small systemic offsets of $\sim 0.05$ mag
in the diagram for $g'$-band images. This offsets may be due to: 1) the effect of CCD QE curve; 2) the
errors of magnitude zero points. However, the CCD QE curve on $g'$ filter is fair flat, and the effect of QE
should be cancelled out in the resulting $g'$ magnitude. Hence, the offset is most likely a zero point
error.

In order to check it out further, we plot the synthetic and observed $(g'-r')$ vs $(u'-g')$ for SDSS filters in
Figure 6. We can see that these plots are very consistent with those in Figure 3. It indicates that the
effect of CCD QE is very tiny. Therefore, we conclude that the offsets between the synthetic and observed
colors are caused by the zero point errors.

We do NOT use SDSS $u'$ and $g'$ magnitudes to do the plots in Figure 5. The reason is that the stars
selected from the $R$-band images are too dim to acquire accurate magnitudes from SDSS catalogues. However,
there are no visible offsets between the observed and synthetic colors after we corrected the errors of the
$g'$-band zero points.

\subsection{Limiting magnitudes}
The sensitivity of the mosaics was tested using a Monte-Carlo procedure. Large numbers of fake point-like
sources are added to the images and then we try to recover them using the same parameters as those used to generate
the real image catalogues. The positions of the artificial sources are assigned randomly apart from the
requirement that they are at least one aperture diameter away from the detected sources.

\begin{figure}[pbht]
\begin{center} 
\includegraphics[width=0.8\textwidth,bb=21 7 496 344]{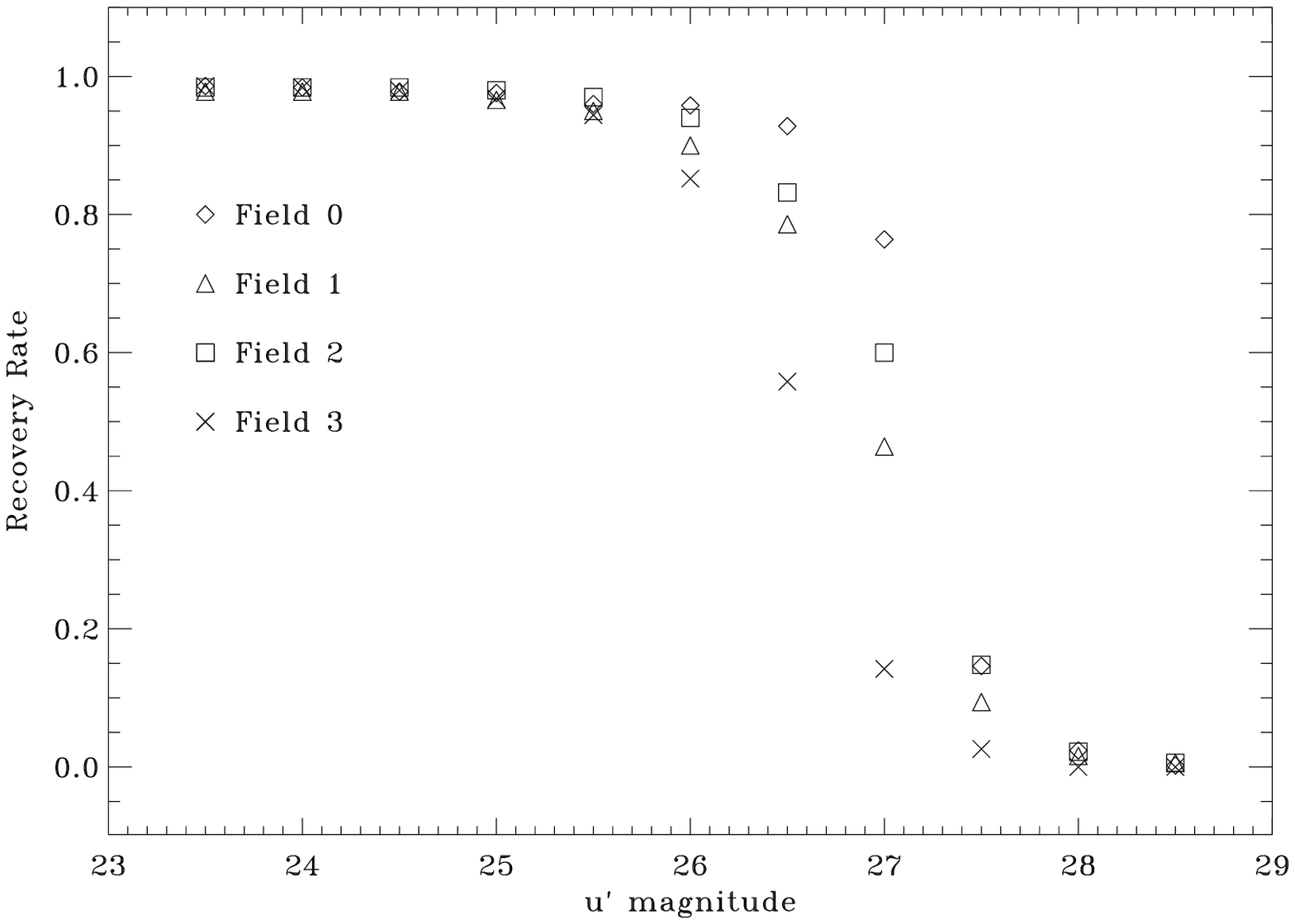} 
\caption{Recovery rates plotted versus magnitudes for the
$u'$-band images. }
\end{center}
\end{figure}
\begin{figure}[pbht]
\begin{center} 
\includegraphics[width=0.77\textwidth,bb=21 7 496 344]{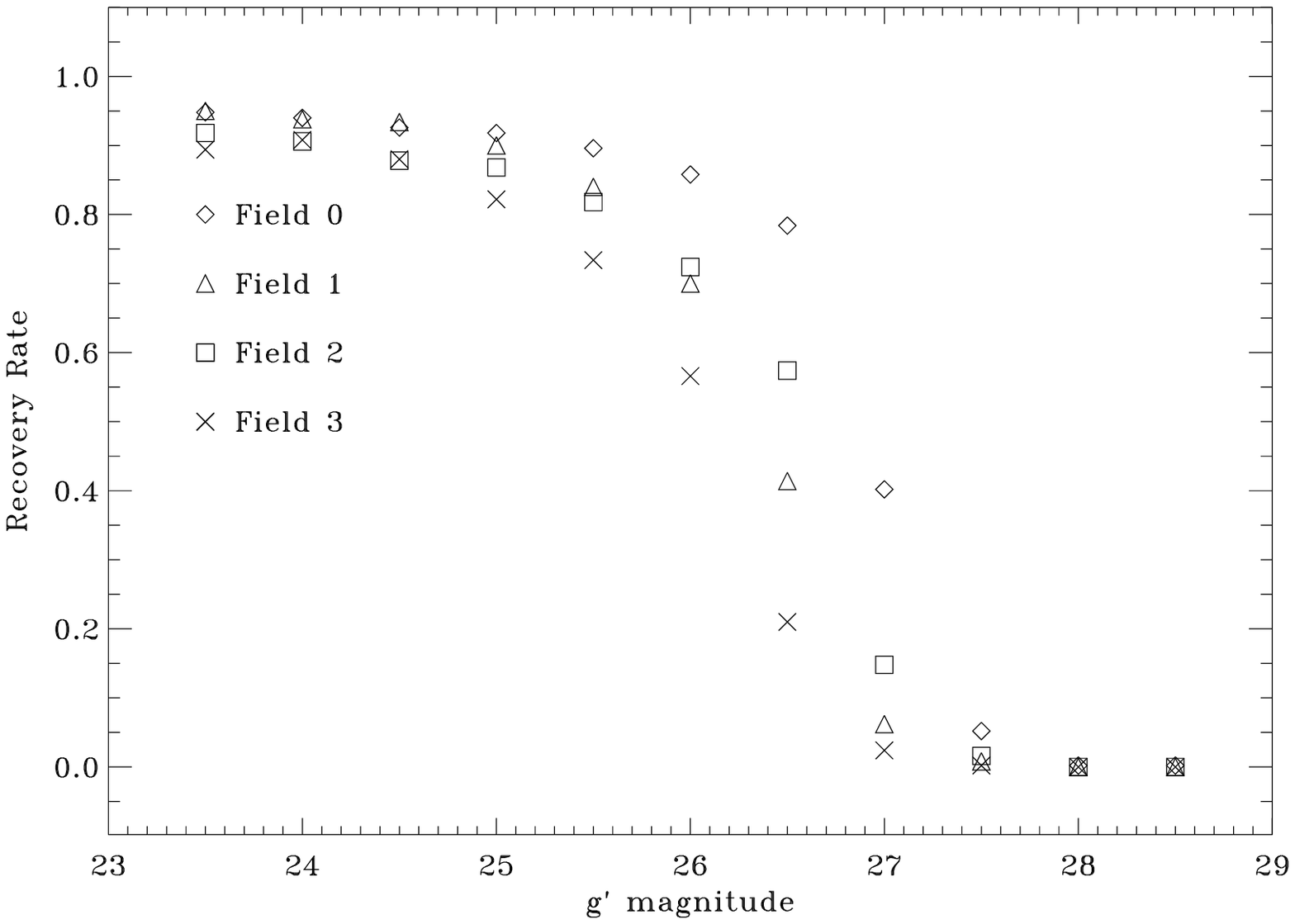} 
\caption{Same as figure 7, for $g'$-band images.}
\end{center}
\end{figure}

In figure 7 we show the recovery rates plotted versus magnitudes for the four $u'$-band images. We can see
that the image of field 0 is the deepest, while field 3 the shallowest. The limiting magnitudes of these 4
fields are $\sim$26.8, 26.6, 26.5 and 26.1 mag, by using 80\% completeness level as the criterion. We'll
find that this results are consistent with those given by the galaxy number counts.

Figure 8 shows the simulation results for $g'$-band images. The scatter, even at the bright end, is much
larger than the other two band images. The large variation of seeing is likely responsible. The limiting
magnitudes are $\sim$26.7, 25.7, 25.8, 25.2 mag.

\begin{figure}[pbht]
\begin{center} 
\includegraphics[width=0.77\textwidth,bb=21 7 496 344]{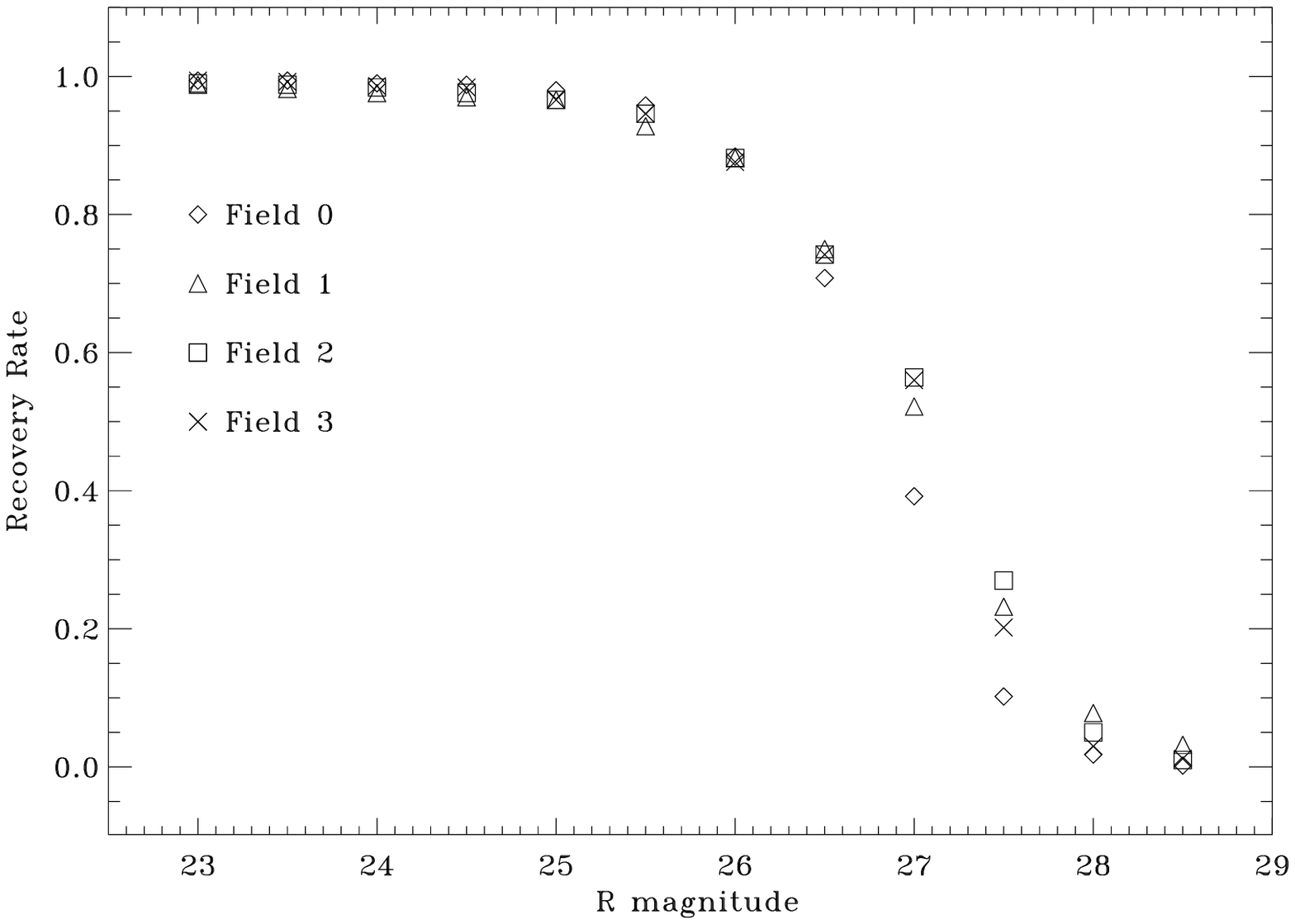}
\caption{Same as figure 7, but for $R$-band images.}
\end{center}
\end{figure}

The $R$-band results are shown in figure 9. As we know, this band has the best image qualities, reflected by
the small scatter, high recovery rate. The limiting magnitudes are $\sim$ 26.2, 26.4, 26.3 and 26.3 mag.

\section{Results and Analysis}
\label{sect:resul} 
\subsection{Galaxy number counts} 

Galaxy number count is a useful tool for
obtaining the depth and homogeneity of the survey, also for constructing galaxy evolution models. In the
following, we present the galaxy number counts in the $u'$-, $g'$- and $R$-bands. The number counts are
constructed for the deep central field in the four subfields first, and only tiny variances among these four
fields are found. Then we take the mean values as the resulting counts. The total areas are 0.25 deg$^2$,
0.10 deg$^2$ and 0.20 deg$^2$, for the $u'$-, $g'$- and $R$-bands, respectively. When doing the counts, no
dereddening were applied.

\subsubsection{Star-galaxy separation}

Although {\scriptsize SExtractor} itself provides a parameter, CLASS\_STAR [with values between 1 (starlike
object) and 0], to separate starlike sources from extended sources, we use the half-light radius vs
magnitude plot (as shown in figure 2) to separate stars from galaxies. This allows us to weed out saturated
sources simultaneously. But this selection criterion doesn't work when sources fainter than 23.5 mag.
However, number counts tend to be dominated by galaxies at the faint end. Therefore, we do not attempt to
separate stars from galaxies at these flux levels.

\subsubsection{$u'$-band counts}

\begin{table}[pbht]
\begin{center}
\caption[]{Differential galaxy counts from the EGS field in $u'$-band.}
\label{table2}
\begin{tabular}{cccc}
\hline\noalign{\smallskip}
Magnitude & & Area & $\log N_{u'}$\\
(AB) &Number of Galaxies&(deg$^2$)& (0.5 mag$^{-1}$ deg$^2$)\\
\hline\noalign{\smallskip}
19.75 & 11 & 0.25 & 1.642\\
20.25 & 15 & 0.25 & 1.777\\
20.75 & 43 & 0.25 & 2.234\\
21.25 & 73 & 0.25 & 2.464\\
21.75 & 143 & 0.25 & 2.756\\
22.25 & 261 & 0.25 & 3.017\\
22.75 & 523 & 0.25 & 3.319\\
23.25 & 1093 & 0.25 & 3.639\\
23.75 & 2151 & 0.25 & 3.933\\
24.25 & 3454 & 0.25 & 4.139\\
24.75 & 5539 & 0.25 & 4.344\\
25.25 & 7490 & 0.25 & 4.475\\
25.75 & 9032 & 0.25 & 4.556\\
26.25 & 9130 & 0.25 & 4.561\\
26.75 & 7525 & 0.25 & 4.477\\
27.25 & 4438 & 0.25 & 4.248\\
27.75 & 1461 & 0.25 & 3.765\\
28.25 & 188 & 0.25 & 2.875\\
28.75 & 36 & 0.25 & 2.157\\
\hline\noalign{\smallskip}
\end{tabular}
\end{center}
\end{table}

\begin{figure}[pbht]
\begin{center} 
\includegraphics[width=\textwidth,bb=18 87 488 605]{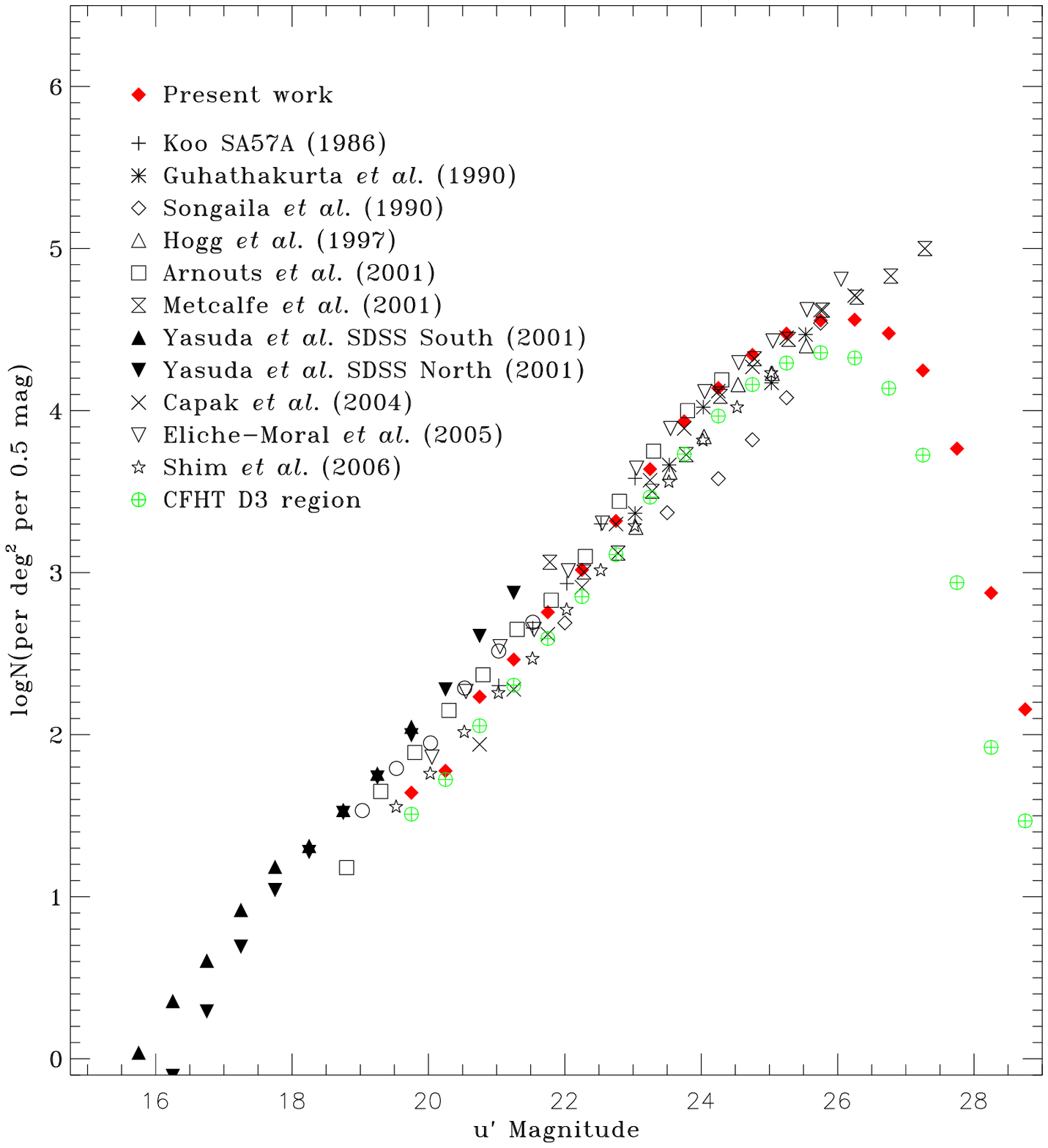}
\caption{$u'$-band differential counts compilation, magnitudes are in AB system.}
\end{center}
\end{figure}

The differential $u'$-band galaxy counts are plotted in figure 10, with a compilation of published galaxy
counts at close bands, eg Johnson/Kron $U$, CFHT $u^*$. All magnitudes have been converted to the SDSS AB
magnitudes, using the conversion relation of galaxy colors from Fukugita et al. (1995). We can see that our
results are very consistent with the other counts, except for that of Songaila et al. (1990), which has
relative lower values than all of the others. Meanwhile, the depth of the survey is deeper than that of CFHT. At the bright end ($u' \leq 20.5$), our counts have a relative
lower value than the others. This is believed due to the small number of galaxies detected in the field. A
small divergency of the number of detected bright galaxies can result in a large variance of the mean
counts. Within the range $21.75 \leq u' \leq 25.75$, the counts have a slope of d(log $N$)/d$m_{u'}
\sim 0.48$. As shown in the figure, our $u'$-band imaging is one of the deepest ground-based observations so
far. Table 2 gives our $u'$-band counts.

\subsubsection{$g'$-band counts}

\begin{table}[pbht]
\begin{center}
\caption[]{Differential galaxy counts from the EGS field in $g'$-band}
\label{table3}
\begin{tabular}{cccc}
\hline\noalign{\smallskip}
Magnitude & & Area & $\log N_{g'}$\\
(AB) &Number of Galaxies&(deg$^2$)& (0.5 mag$^{-1}$ deg$^2$)\\
\hline\noalign{\smallskip}
19.25 & 14 & 0.10 & 2.105\\
19.75 & 18 & 0.10 & 2.246\\
20.25 & 29 & 0.10 & 2.475\\
20.75 & 54 & 0.10 & 2.735\\
21.25 & 92 & 0.10 & 2.957\\
21.75 & 151 & 0.10 & 3.150\\
22.25 & 297 & 0.10 & 3.457\\
22.75 & 404 & 0.10 & 3.599\\
23.25 & 789 & 0.10 & 3.902\\
23.75 & 1327 & 0.10 & 4.112\\
24.25 & 1889 & 0.10 & 4.271\\
24.75 & 2685 & 0.10 & 4.424\\
25.25 & 3522 & 0.10 & 4.537\\
25.75 & 4135 & 0.10 & 4.603\\
26.25 & 4067 & 0.10 & 4.598\\
26.75 & 3339 & 0.10 & 4.501\\
27.25 & 1698 & 0.10 & 4.205\\
27.75 & 414 & 0.10 & 3.600\\
28.25 & 48 & 0.10 & 2.674\\
\hline\noalign{\smallskip}
\end{tabular}
\end{center}
\end{table}

We show the $g'$-band counts in figure 11, compilation with the results of $g'$-band from SDSS (Yasuda et
al. 2001), $B$-band from the ground-based observations, and $F450$-band from HDF north and south fields. It
can be seen that our observation is deeper than most of the ground-based observations, but much shallower
than the Hubble Space Telescope obsertions. The average slope d(log N)/dm is 0.4 for $20 \leq g'
\leq 25.5$. The shallower slope fainter than 25.5 is caused by the incompleteness. However, it is also
within $3\sigma$ at $\sim 26.0$ mag. Our $g'$-band counts are listed in Table 3.

\subsubsection{$R$-band counts}
\begin{table}[pbht]
\begin{center}
\caption[]{Differential galaxy counts from the EGS field in $R$-band}
\label{table4}
\begin{tabular}{cccc}
\hline\noalign{\smallskip}
Magnitude & & Area & $\log N_{R}$\\
(AB) &Number of Galaxies&(deg$^2$)& (0.5 mag$^{-1}$ deg$^2$)\\
\hline\noalign{\smallskip}
19.25 & 35 & 0.20 & 2.248\\
19.75 & 86 & 0.20 & 2.639\\
20.25 & 146 & 0.20 & 2.869\\
20.75 & 228 & 0.20 & 3.062\\
21.25 & 424 & 0.20 & 3.332\\
21.75 & 585 & 0.20 & 3.471\\
22.25 & 961 & 0.20 & 3.687\\
22.75 & 1399 & 0.20 & 3.850\\
23.25 & 2095 & 0.20 & 4.025\\
23.75 & 3204 & 0.20 & 4.210\\
24.25 & 4884 & 0.20 & 4.393\\
24.75 & 6982 & 0.20 & 4.548\\
25.25 & 10086 & 0.20 & 4.708\\
25.75 & 13644 & 0.20 & 4.839\\
26.25 & 18712 & 0.20 & 4.976\\
26.75 & 23558 & 0.20 & 5.076\\
27.25 & 28233 & 0.20 & 5.155\\
27.75 & 21537 & 0.20 & 5.037\\
28.25 & 5773 & 0.20 & 4.466\\
\hline\noalign{\smallskip}
\end{tabular}
\end{center}
\end{table}

In figure 12 we plot the $R$-band counts. The SDSS $r'$ counts (Yasuda et al. 2001) and $F606$ counts on HDF
north and south fields (Metcalfe et al. 2001) are also included, as well as other ground-based $R$-band
observations. Our result agrees with the others very well. We can see that our observation is almost the
deepest ground-based $R$-band observations up to now (up to $\sim 26.5$ mag), which is much deeper than CFHT's, and only the counts presented
by Kashikawa et al. (2004) can be comparable with ours. The counts continue to increase, with an average
slope of 0.36 for $20\leq R \leq 26$ mag, but with evidence of a change to a somewhat shallower
slope faintward of this. This apparent turnover is due to incompleteness. The results for the $R$-band are
given in Table 4.
\clearpage
\newpage
\begin{figure}
\begin{center} 
\includegraphics[width=\textwidth,bb=28 5 496 600]{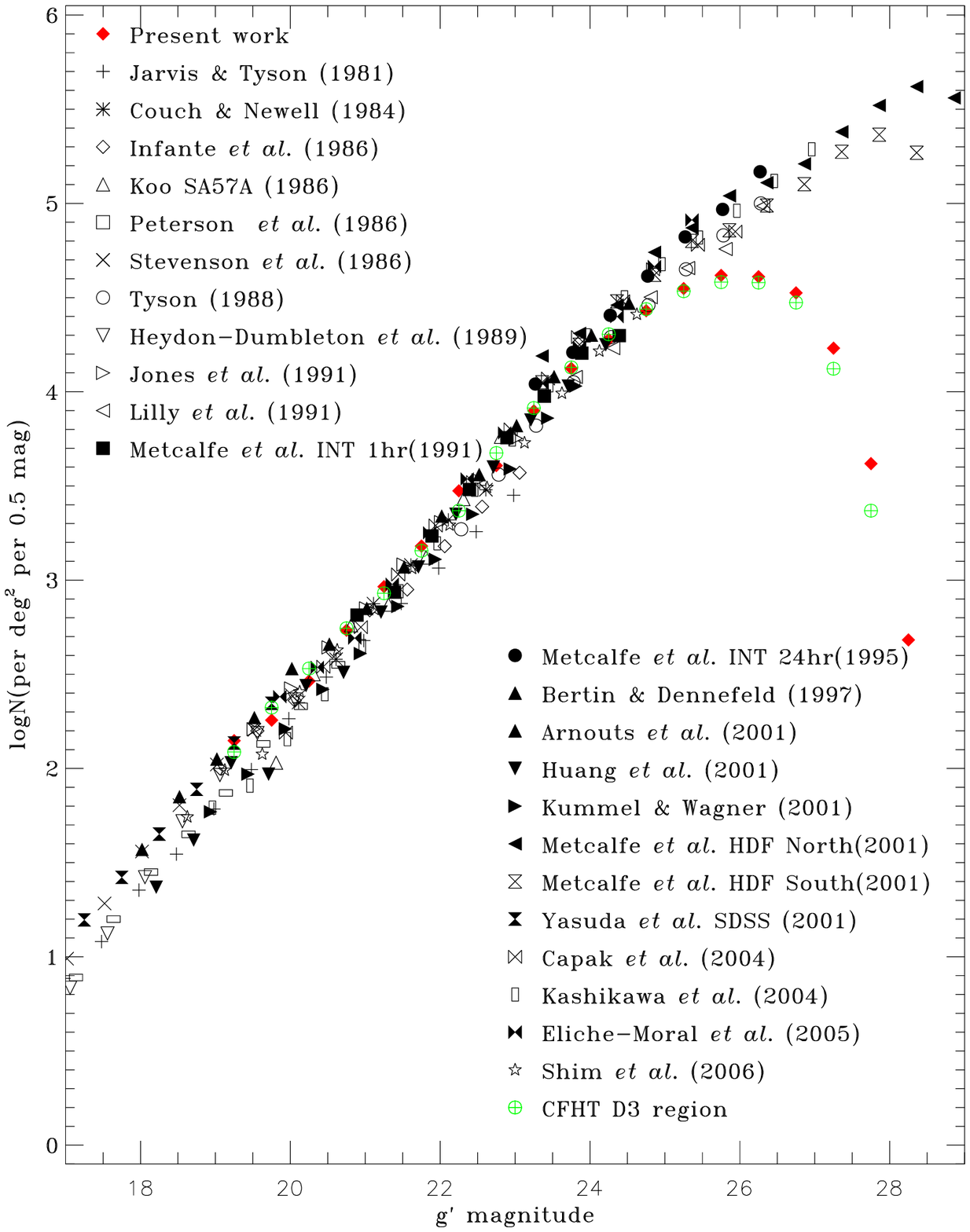}
\caption{$g'$-band differential counts compilation, magnitudes are in AB system.}
\end{center}
\end{figure}
\clearpage
\newpage
\begin{figure}
\begin{center} 
\includegraphics[width=\textwidth,bb=17 58 488 587]{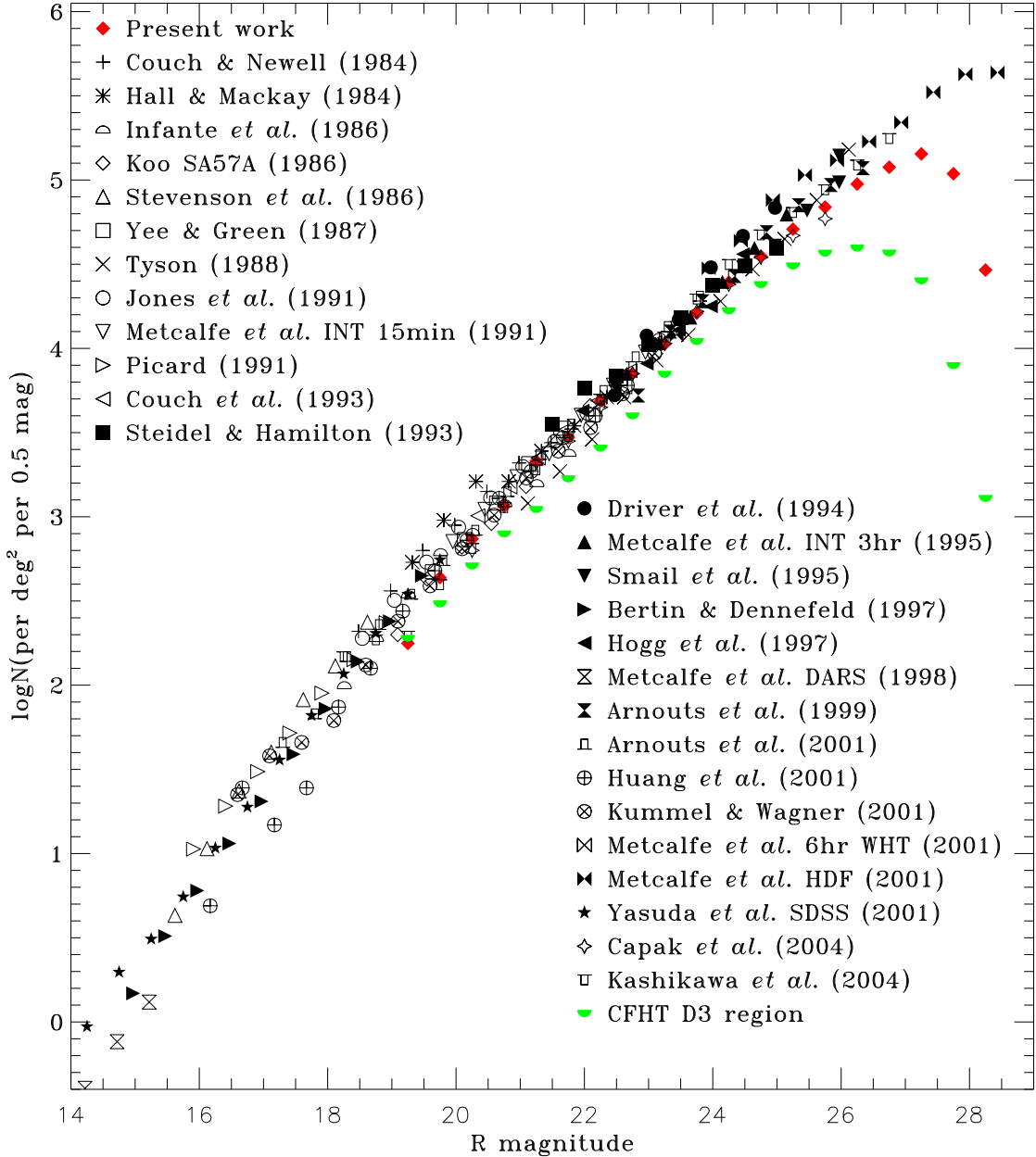}
\caption{$R$-band differential counts compilation, magnitudes are in AB system. We can see that our depth at this band is much deeper than the CFHT's.}
\end{center}
\end{figure}

\subsection{$u'g'R$ Photometric Selection of Lyman Break Galaxies}

\begin{figure}[pbht]
\begin{center} 
\includegraphics[width=0.8\textwidth,bb=50 0 408 360]{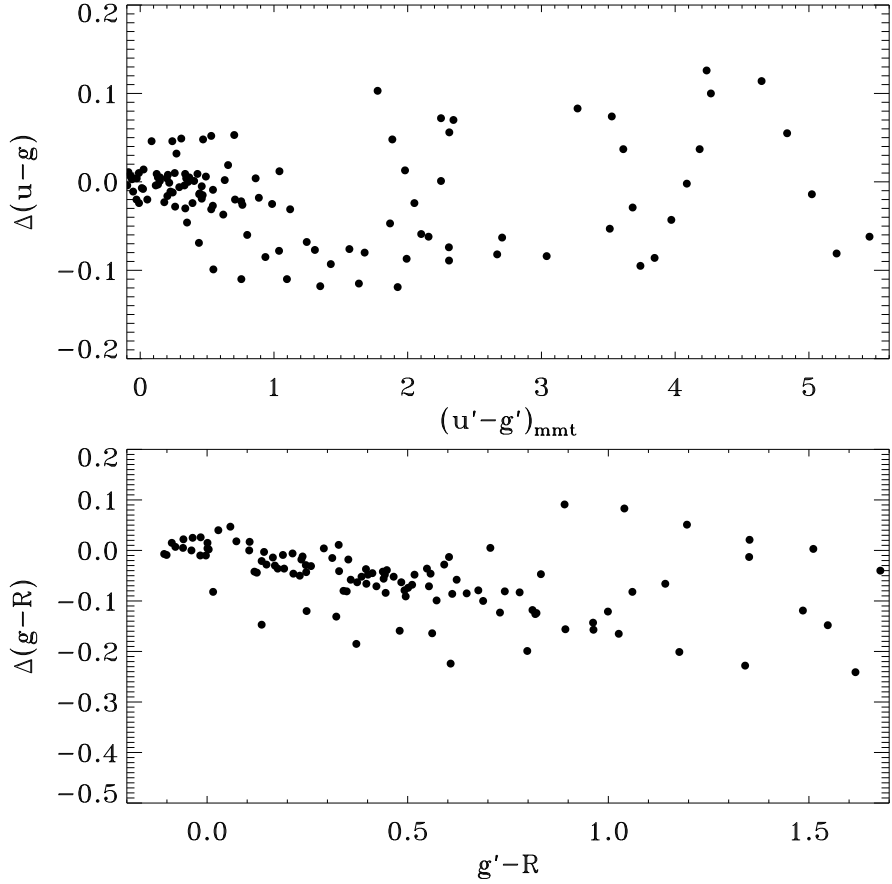}
\caption{The dependence of color difference between our filter system and filter system used by Steidel, on
our color. We can see a strong correlation in the bottom panel.}
\end{center}
\end{figure}

\begin{figure}[pbht]
\begin{center} 
\includegraphics[width=0.8\textwidth,bb=24 50 497 512]{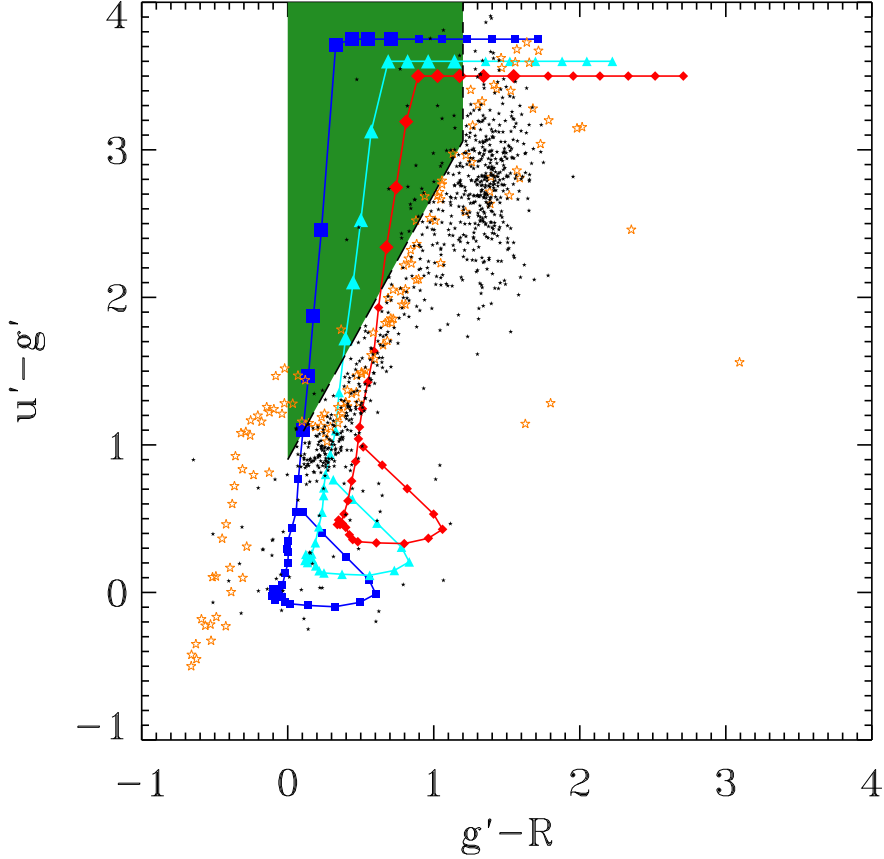}
\caption{$(u'-g')$ vs. $(g'-R)$ colors in our filter sysmtem (AB magnitudes) of a star-forming galaxy with a  constant star formation rate in the targeted redshift range, for three assumed values of internal extinction [E(B-V)=0, 0.15, 0.30 using the Calzetti et al. 2000 prescription, for squares, triangles, and diamonds, respectively], with points corresponding to intervals of $\Delta z=0.1$. The large points on each curve correspond to galaxies in the redshift interval $2.7 \leq  z \leq  3.4$. The $(u'-g')$ colors have been truncated for clarity. The expected location of the stellar locus, based on the  stellar spectra library HILIB (Pickles 1998), is shown with orange stars. The observational stellar colors are shown with black stars. The green shaded region shows our adopted color selection criteria. }
\end{center}
\end{figure}

\begin{figure}[pbht]
\begin{center} 
\includegraphics[width=0.8\textwidth,bb=12 5 492 345]{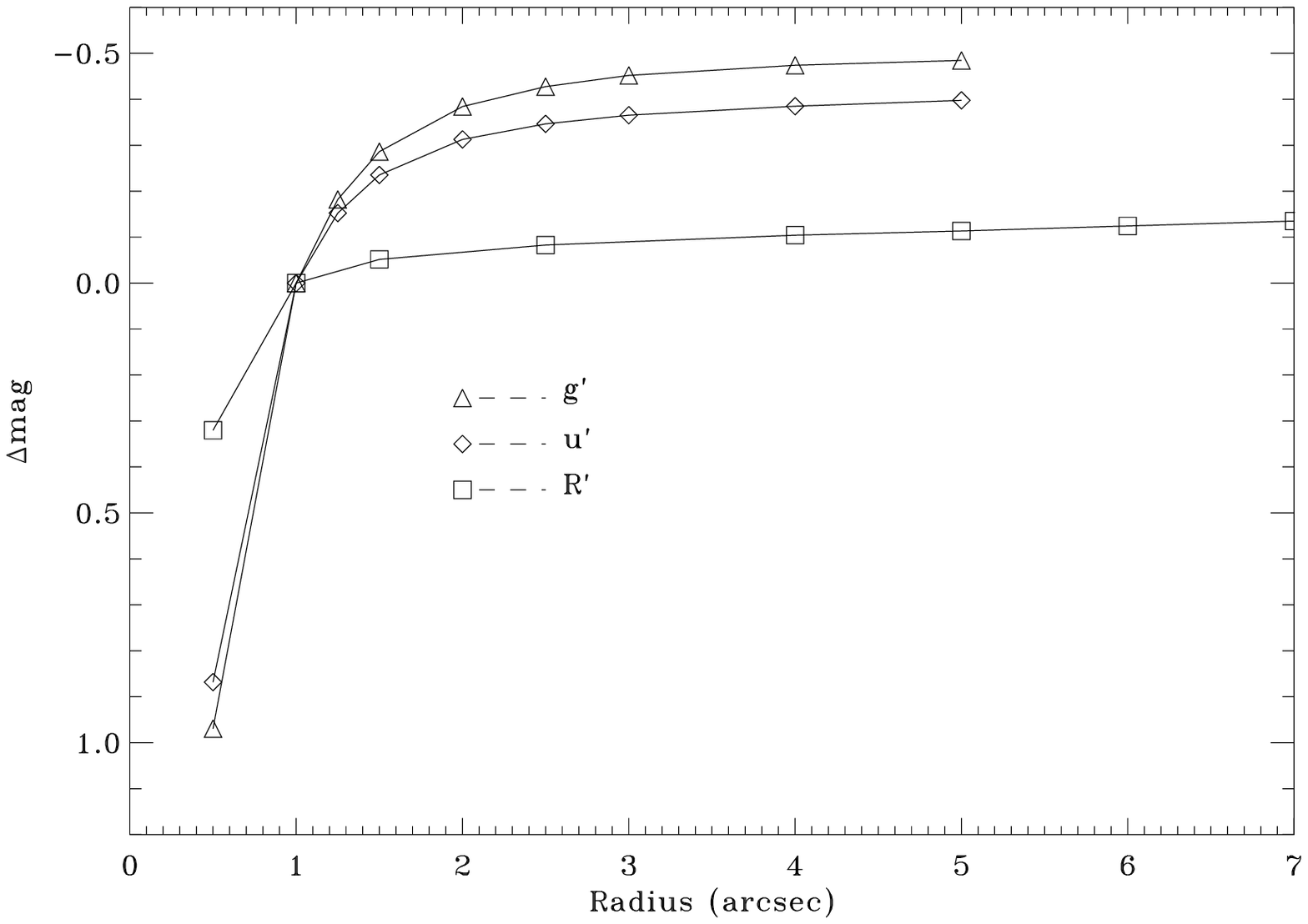}
\caption{The curve of growth of stars for $u'$-,$g'$-and $R$-band images in Field 1.}
\end{center}
\end{figure}

Various filter sets have yielded success at finding LBGs, such as $U_nG\mathcal{R}$ (Steidel et al. 1996b),
$G\mathcal{R}i$ ( Steidel et al. 1999), $BRI$ (Gawiser et al. 2001; Prochaska et al. 2002), SDSS
$u'g'r'i'z'$ (Bentz et al. 2004), $u'BVRI$ (Cooke et al. 2005), and $BVR$ (Gawiser et al. 2006). Figure 13
shows the difference of expected colors of model star-forming galaxies between our filter system and Steidel's
filter system, with a variety of assumed redshifts and reddening by dust. There is no obvious difference between the
$(u'-g')$ colors at $0.\leq(u'-g')\leq1.0$. However, there exists a large discrepancy when $(u'-g')>1$, which is the color range for almost all of the LBG candidates. For the case of $(g'-R)$ color,
there exists a strong correlation between color difference, $\Delta (g'-R)=(g'-R)-(G-\mathcal{R})$ and the
$(g'-R)$ color. We use a minimum $\chi^2$ fit to obtain: $\Delta (g'-R)=-0.01-0.16 (g'-R)$.  However, it is too complicated
to transform Steidel et al. criteria to our filter system. Hence, we generated our own color selection criteria. In figure 14 we show
the expected colors (through our filter system) versus redshift of model star-forming galaxies with a variety of 
assumed reddening by dust. We also plotted the expected and the observational star colors and our adopted color selection criteria of
\[ \begin{array}
{*{20}c} {u' - g' \ge 1.8(g' - R) +
0.9,} \\ { 0.0 \le g' - R \le 1.2,} \\ {19.0 \le R \le 25.0,} \\ 
\end{array} \] 
where all magnitudes are in AB.  Since our observations ($g'$-band) are 
not as deep as Steidel's, we select LBGs up to $R=25.0$ mag.  In order to avoid the contamination from the upper branch of the stellar distribution at red colors, we shifted our criteria to avoid this region as it also contains dim dwarf stars with correspondingly large errors in color, which are the primary expected source of interlopers. The region of the Steidel et al. (2003) criteria avoided by this shift had the highest fraction of interlopers.

In order to obtain accurate colors of the sources, we need measure the flux within an aperture with the same
size, assuming that the PSFs are nearly same. However, the PSFs are much different for our $u'g'R$-band images since
they were obtained by using different detectors and under much different weather conditions (see Table
1). The seeing of $g'$-band is always almost twice as big as that of $R$-band. In figure 15 we show the
curve of growth for stars in F1 field. We can see that the profiles between $R$-band and $u'$- and
$g'$-bands are much difference. What's more, these profiles are non-Gaussian, and can be well fitted by a
moffat function. Hence, we do not simply use a Gaussian kernel to convolve the $u'$- and $R$-band images to
match $g'$-band images, but to deconvolve the $g'$-band PSF at first, then to convolve the original $u'$-
and $R$-band images using the deconvolved kernels. The detection of objects was performed on the original
rebinned $R$-band image, not the smoothed image, by using the {\scriptsize Source Extractor}'s double-image
mode. Objects were thought to be detected if the number of connected pixels with flux exceeding 3 times the sky $\sigma$
is larger than 4. However, this has no much effect on our results, since our cut off magnitude of $R$
is 25.0 mag. The aperture size we chose is a 3$''$ diameter circular aperture, to match the IRAC aperture size. The total $R$-band magnitude is measured on the original image by using a same size
aperture, which can include almost all light of a point source, from the curve of growth shown in figure 15.

\begin{table}[pbht]
\begin{center}
\caption[]{LBG candidates selected in these four EGS fields}
\label{table5}
\begin{tabular}{cccc}
\hline\noalign{\smallskip}
 & Area & &Surface density\\
 Field&(arcmin$^2$)&Number & (arcmin$^{-2}$)\\
\hline\noalign{\smallskip}
0 & 240.4 & 195 & 0.8\\
1 & 916.4 & 1020 &1.1 \\
2& 153.0 & 167 & 1.1\\
3 & 327.4 & 260 &0.8 \\
Total&1637.3 &1642&1.0\\
\hline\noalign{\smallskip}
\end{tabular}
\end{center}
\end{table}

Since many sources are undetected in $u'$-band image, objects with $u'$ flux, $f_{u'}$, less than their
$1 \sigma$ flux uncertainty, $\sigma_{u'}$, are assigned an upper limit of $f_{u'}=\sigma_{u'}$, where
$\sigma_{u'}=N^{0.5}_{\rm{pix}} \sigma_{\rm{sky}}$, and $N_{\rm{pix}}$ is the number of pixels in the detection aperture  and $\sigma_{\rm{sky}}$ is the rms pixel-to-pixel
fluctuations in the sky background. This procedure is the same as Steidel et al. (2003). The limit values are used to
generate the $(u'-g')$ colors. Our selection criteria yield 1642 candidates in 1637.3 arcmin$^2$, as listed in Table 5. When doing
the LBG selections, we have trimmed the images of Field 0, Field 2 and Field 3, according to the
signal-to-noise ratio. The average surface density of the candidates is $\sim1.0$ galaxies arcmin$^{-2}$,
which is slightly higher than that of $\sim0.9$ galaxies arcmin$^{-2}$ (at $R_{AB}=25.0$ mag), found by Steidel et al. (2003).
We believe that this discrepancy is mainly caused by the observational depth and/or the cosmic variance.

\begin{figure}[pbht]
\begin{center} 
\includegraphics[width=0.8\textwidth,bb=10 5 490 344]{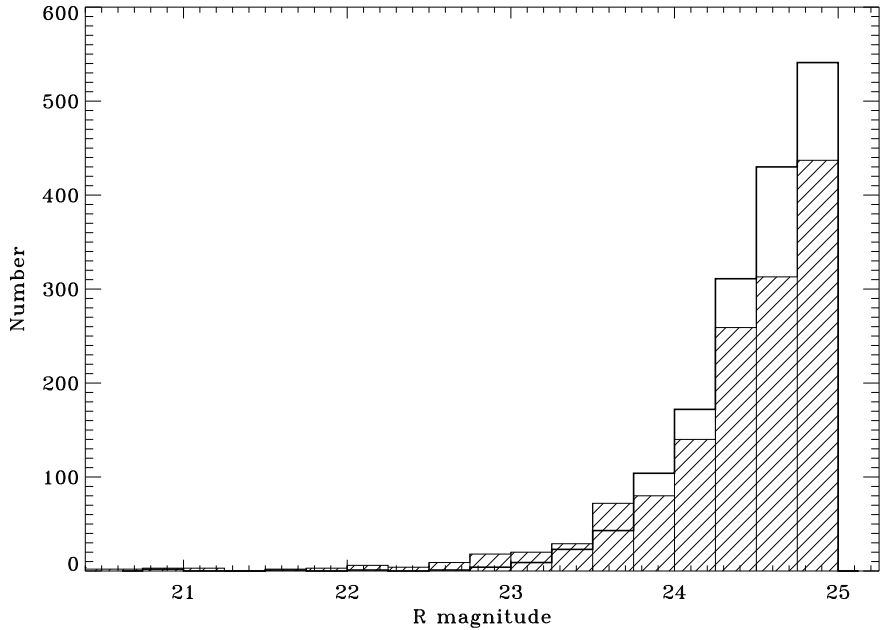}
\caption{Magnitude distributions of the full photometric LBG sample with $19.0 \le R_{AB}\le 25.0$ (thick line). We only plot up to $R \ge 20.5$ since there are no LBGs brighter than 20.5 mag in the sample. It can be seen that this distribution is consistent with that (hashed histogram) in Steidel et al. (2003).}
\end{center}
\end{figure}

Figure 16 shows the magnitude distribution of the full photometric selected LBG sample\footnotemark{}\footnotetext{This catalogue will be updated since new data were obtained.}. It is very similar
to that shown in Steidel et al. (2003). Most LBG candidates are dim sources, and we found no LBGs brighter than
20.5 mag in this sample. In the next papers of this series, we will present the properties for the LBG sample, such 
as photometric redshift, stellar mass, and the feature of clustering.

\section{Summary}
\label{summary}
In this paper we present the photometry for the Extended Groth Strip $u'g'R$ survey, assembled by using the
MMT/Megacam and the Subaru/Suprime. This papers has detailedly described the photometric calibration,
limiting magnitudes and galaxies number counts. The galaxies number counts are very consistent with previous
results. The number counts show that our $u'$-band observation is one of the deepest ground-based
observations so far, and our $g'$-band observation is deeper than most of the ground-band observations. The
counts also show that our $R$-band observation reaches $26.2 \sim 26.5$ mag, which allows our observation to
be the deepest ground-based observations up to now. These also mean that our observations are deep enough to
study the high redshift and consequently dim objects.

We also present a photometrically selected LBG sample, as the first scientific result. The high quality of
the $R$-band images may resolve the high redshift objects and allow us to study their structures. In the
next papers of this series, we will present some other scientific results for the LBG sample.

\begin{acknowledgements} 
The observations reported here were obtained in part at the MMT Observatory, a facility
operated jointly by the Smithsonian Institution and the University of Arizona. YHZHAO would like to gratefully
acknowledge the financial supports from China Scholarship Council (CSC), Jiangsu Planned Projects for Postdoctoral Research Funds (No. 0802031C) and K.C.Wong Education Foundation, Hong Kong.
\end{acknowledgements}

\end{document}